\documentclass[10pt,aps,twocolumn, pra, footinbib,superscriptaddress]{revtex4-1}
\usepackage{graphicx}
\usepackage{psfrag}
\usepackage{amsmath,amssymb}
\usepackage{colordvi}
\usepackage{color}
\usepackage{hyperref}
\hypersetup{
    colorlinks=true, 
    linktoc=all,     
    linkcolor=blue,  
    urlcolor=blue,
    citecolor=red,
}

\begin{document}

\title{A strongly interacting high partial wave Bose gas}
\author{Juan Yao}
\affiliation{Institute for Advanced Study, Tsinghua University, Beijing, 100084, China}
\author{Ran Qi}
\affiliation{Department of Physics, Renmin University of China, Beijing, 100872, P. R. China}
\author{Pengfei Zhang}
\affiliation{Institute for Advanced Study, Tsinghua University, Beijing, 100084, China}

\date{\today}

\begin{abstract}
Motivated by recent experimental progress, we make an investigation of $p$- and $d$-wave resonant Bose gas. An explanation of the  Nozi{\`e}res and Schmitt-Rink (NSR) scheme in terms of two-channel model is provided. Different from the $s$-wave case, high partial wave interaction supports a quasi-bound state in the weak coupling regime. Within the NSR approximation, we study the equation of state, critical temperature and particle population distributions. We clarify the effect of the quasi-bound state on the phase diagram and the dimer production. A multi-critical point where normal phase, atomic superfluid phase and molecular superfluid phase meet is predicted within the phase diagram. We also show the occurrence of a resonant conversion between solitary atoms and dimers when temperature $k_BT$ approximates the quasi-bound energy.
\end{abstract}
\maketitle
\section{Introduction} 
In ultracold atomic system, Feshbach resonance as a powerful technique has been used to tune the interactions of both fermionic and bosonic gases. Up to now, many experimental and theoretical studies have been focused on $s$-wave resonant system. Recently, high partial wave interactions for its specific properties have attracted more and more attentions. Such as the measurement of contacts in a $p$-wave interacting $^{40}$K gas~\cite{Luciuk2016} and the relevant theoretical discussions~\cite{Yu2015, Yoshida2015, Mingyuan2016, Cui2016, Peng2016a}.
Many efforts have been spent on realization of the high partial wave Feshbach resonance in Fermi gas~\cite{Regal2003, Zhang2004, Gunter2005, Schunck2005}. Compared with Fermi gas which benefits from the Pauli exclusion principle, Bose gas suffers great loss near resonance~\cite{Rem2013} and might undergo instability with attractive interaction~\cite{Stoof1994}.  As a result, the realization of high partial wave resonance in Bose gas is even more challenging. The exciting new progress in this direction is the observation of broad $d$-wave shape resonance in $^{41}$K gas~\cite{dwaveExp} and together with $p$-wave shape resonance in the mixture of $^{85}$Rb-$^{87}$Rb \cite{Cui2017, Dong2016a}. 

Motivated by these progresses, we make an investigation on the strongly interacting high partial wave Bose gas in terms of two-channel model. 
High partial wave interaction contains a centrifugal barrier which supports a quasi-bound state in the weakly coupled regime. The appearance of the quasi-bound state will lead to various new novel phenomena. In this report, we will mainly clarify how the existence of quasi-bound state has an effect on the superfluid phase diagram and the particle population distributions. An transition from solitary atomic superfluid to molecular superfluid is achieved. Within the phase diagram, we predict the existence of  a multi-critical point where normal phase, atomic superfluid phase and molecular superfluid phase meet. Generally, low temperature will benefit the dimer production. In the weak coupling side, we will present a abnormal behavior of the particle population distribution in terms of temperature. Then the appearance of the turning point in the particle population distribution can be explained by  occurrence of a resonance conversion between solitary atoms and molecules when $k_BT$ approximates the quasi-bound state energy.

\section{The Model} For a $d$-wave interacting spinless Bose system, as proposed in Ref.~\cite{Pengfei2016}, the Lagrangian density of the effective field theory can be casted as  
\begin{equation}
\begin{aligned}
&\mathcal{L}=\psi^\dagger(i\partial_t+\frac{\nabla^2}{2M})\psi
+\sum_{m}\frac{\bar{g}_m}{\sqrt{2}}\left[d_{lm}^\dagger \mathcal{Y}_{m} +\text{h.c.}\right] \\
&-\sum_{m}d_{lm}^\dagger \left[i\partial_t+\frac{\nabla^2}{4M}+\bar{z}_m\left(i\partial_t+\frac{\nabla^2}{4M}\right)^2-\bar{\nu}_m\right]d_{lm} , 
\end{aligned}
\label{EqLag}
\end{equation}
where the field operator $\psi $ is the annihilation operator for spinless Bosons. $d_{lm}$ is the dimer field with azimuthal quantum number m and $l=2$ for the $d$-wave interaction. Then the first term corresponds to the free atoms with $M$ is the mass of single atom and $\hbar$ will be set to $1$ throughout the report. The third term corresponds to the free dimers with $\bar{\nu}_m$ being the detuning and $\bar{z}_m$ the bare coupling constant. The conversion between the atoms and dimers is described by the second term with $\mathcal{Y}_m $ defined as 
\begin{equation}
\begin{aligned}
\mathcal{Y}_m=
\sum_{a,b}\frac{C_{a,b}^m}{4} \Big\{
[\partial_a\psi({\bf r}_1)][\partial_b\psi({\bf r}_2)] -[\partial_a\partial_b\psi({\bf r}_1)]\psi({\bf r}_2)& 
\\ +[\partial_b\psi({\bf r}_1)][\partial_a\psi({\bf r}_2)]
-\psi({\bf r}_1) [\partial_a\partial_b\psi({\bf r}_2)] \Big\}&.
\end{aligned}
\end{equation}
where $a$ and $b$ take values $x,y,z$.  $C_{ab}^m$ are the Clebsch-Gordon coefficients when transforming $k_ak_b/k^2$ to the spherical harmonics, which satisfies $\sum_{a,b}C_{a,b}^mk_ak_b/k^2=\sqrt{4\pi}Y_{2m}(\hat{k})$. In terms of $a_{\bf k}$ and $b_{m,{\bf q}}$ (the Fourier transformation of the operators $\psi $ and $d_{lm} $), the atom-dimer coupling in the Lagrangian $\mathcal{L}=\int d{\bf r} \mathcal{L} $ takes the form 
\begin{equation}
\begin{aligned}
\mathcal{L}&_{ad}=\\
&\sum_{m}\sum_{\bf q, k}\sqrt{\frac{2\pi}{V}}\bar{g}_m\left[
k^2Y_{2m}(\hat{k})b_{m,{\bf q}}^\dagger a_{\frac{\bf q}{2}+{\bf k}} a_{\frac{\bf q}{2}-{\bf k}}
+\text{h.c.}
\right],
\end{aligned}
\end{equation}
where $V$ is the total volume and $\bar{g}_m$ is the bare interaction strength.
In order to renormalize the bare coupling parameters $\bar{g}_m$, $\bar{\nu}_m$, $\bar{z}_m$, we construct an effective low energy field theory and expand the $d$-wave phase shift up to $k^4$ with $k^5\cot\delta_m(k)=-1/D_m-k^2/v_m-k^4/R_m$. The super volume $D_m$, the effective volume $v_m$ and the effective range $R_m$ are the minimal set of parameters that are needed to renormalize Eq.~\eqref{EqLag}. Considering the scattering between two Bosons with total momentum zero and relative momentum $2k\hat{k}$, the associated $T$-matrix is derived as 
\begin{equation}
\begin{aligned}
T(k\hat{k}^\prime,k\hat{k},&E)= \\
&\frac{\frac{8\pi\bar{g}_m^2}{V}k^4Y_{2m}^*(\hat{k}^{\prime})Y_{2m}(\hat{k})}{-E-\bar{z}_mE^2+\bar{\nu}_m-\frac{1}{V}\sum_{{\bf k}_1}\frac{4\pi\bar{g}_m^2k_1^4|Y_{2m}(\hat{k}_1)|^2}{E-\frac{k_1^2}{M}}},
\end{aligned}
\end{equation}
where the relative outgoing momentum is denoted as $2k\hat{k}^\prime$. Matching $T(k\hat{k}^\prime,k\hat{k},E=\frac{k^2}{M}+i0^+)$ with the expansion of phase shift $\cot\delta_m(k)$ in the limit of $k\to0$, we obtain the following renormalization relations:
\begin{equation}
\begin{aligned}
\frac{\bar{\nu}_m}{\bar{g}_m^2}&=\frac{M}{4\pi}D_m^{-1}-\frac{M}{V}\sum_{\bf k} k^2,  \\
\frac{1}{\bar{g}_m^2}&=-\frac{M^2}{4\pi}v_m^{-1}+\frac{M^2}{V}\sum_{\bf k} 1, \\
\frac{\bar{z}_m}{\bar{g}_m^2}&=-\frac{M^3}{4\pi}R_m^{-1}+\frac{M^3}{V}\sum_{\bf k} k^{-2}.
\end{aligned}
\label{EqNorm}
\end{equation}

Similar to the $d$-wave resonance, a $p$-wave interacting Feshbach resonance in a two-component Bose gas can be described by
\begin{equation}
\begin{aligned}
\hat{H}
=&\sum_{{\bf k,\sigma}}\left(\frac{k^2}{2M}-\mu\right)\hat{a}_{{\bf k},\sigma}^\dagger\hat{a}_{{\bf k},\sigma}\\
&+\sum_{{\bf q},m}\left(\frac{q^2}{4M}-\bar{\nu}_m-2\mu\right)\hat{b}_{{\bf q},m}^\dagger\hat{b}_{{\bf q},m}\\
&+\frac{1}{\sqrt{V}}\sum_{{\bf k},{\bf q},m}\bar{g}_m kY_{1m}(\hat{{\bf k}})\hat{b}_{{\bf q},m}^\dagger\hat{a}_{\frac{{\bf q}}{2}+{\bf k},\uparrow}\hat{a}_{\frac{{\bf q}}{2}-{\bf k},\downarrow}+{\rm h.c.},
\end{aligned}
\end{equation}
where $\sigma=\uparrow$ and $\downarrow$ denoting the pseudo-spin. Different from the $d$-wave case, scattering can only occur between two atoms with opposite spin. For the $p$-wave case, the low energy expansion up to $k^2$ involves only two parameters with $k^3\cot\delta_m(k)=-1/v_m-k^2/R_m$ which result in the renormalization relations:
\begin{equation}
\begin{aligned}
\frac{\bar{\nu}_m}{\bar{g}_m^2}&=\frac{M}{4\pi}v_m^{-1}-\frac{M}{V}\sum_{\bf k} 1,  \\
\frac{1}{\bar{g}_m^2}&=\frac{M^2}{4\pi}R_m^{-1}-\frac{M^2}{V}\sum_{\bf k} \frac{1}{k^2}.
\end{aligned}
\end{equation}

\section{Nozi{\`e}res and Schmitt-Rink scheme}  
In this report, we build the formalism in the normal phase. The thermodynamic potential will contain three parts with respect to each term in the total Hamiltonian: $\Omega=\Omega_0^{\rm B}+\Omega_0^{\rm M}+\Omega_{\text{int}}$. Here $\Omega_0^{\rm B}=1/\beta \sum_{\bf k}\ln \left[1-e^{-\beta\xi_k}\right]$ gives the contribution of non-interacting Bosons with $\beta=1/k_{\rm B}T$ and $\xi_k=\frac{k^2}{2M}-\mu$ being the kinetic energy of an atom measured from its chemical potential $\mu$. $\Omega_0^{\rm M}$ gives the contribution from dimers. Within the NSR scheme, the contribution from the interaction term $\Omega_{\text{int}}$ can be evaluated by taking account of all orders of ring diagrams~\cite{Nozires1985}. Finally, $\Omega_{\text{int}}$ is explicitly given by 
\begin{equation}
\begin{aligned}
\Omega_{\rm int} =&\sum_{m, {\bf q}}\int_{-\infty}^{+\infty}\frac{d\omega}{\pi}\frac{1}{e^{\beta\omega}-1}\times \\
&{\rm Im}\big\{\ln \left[ 
1+\bar{g}_m^2\Pi({\bf q},\omega+i0^+)G_0^{\rm M}({\bf q},\omega+i0^+)
\right]\big\},
\end{aligned}
\end{equation}  
in which $\Pi({\bf q},\omega)$ is the pair-pair propagator written as
\begin{equation}
\begin{aligned}
\Pi^d({\bf q},\omega)=\Pi^{l=2}({\bf q},\omega)&=\frac{1}{V}\sum_{\bf k}
\Big\{k^{2l}4\pi|Y_{lm}(\hat{k})|^2 \\
&\times \frac{1+n(\xi_{{\bf k}+{\bf q}/{2}})+n(\xi_{-{\bf k}+{\bf q}/2})}{\xi_{{\bf k}+{\bf q}/2}+\xi_{-{\bf k}+{\bf q}/2}-\omega} \Big\},
\end{aligned}
\end{equation}
where $n(\xi)=1/(e^{\beta\xi}-1)$ is the Bose distribution function and $l=2$ denoting the $d$-wave interaction. For $d$-wave resonance, the free Green's function for the dimer is 
\begin{equation}
\begin{aligned}
(G_0^{\rm M})^{-1}({\bf q},\omega)=&-(\omega-\frac{q^2}{4M}+2\mu)+\bar{\nu}_m \\
&-\bar{z}_m(\omega-\frac{q^2}{4M}+2\mu)^2.
\end{aligned}
\end{equation}
To get a more concise formulation, we write $\Omega_{\rm int}$ into two parts:
\begin{equation}
\begin{aligned}
\Omega_{\rm int} =&\sum_{m, {\bf q}}\int_{-\infty}^{+\infty}\frac{d\omega}{\pi}\frac{1}{e^{\beta\omega}-1}  {\rm Im} \Big\{
\ln 
G_0^{\rm M}({\bf q},\omega+i0^+)+ \\
 &\ln \left[ 
\left(G_0^{{\rm M}}\right)^{-1}({\bf q},\omega+i0^+)+\bar{g}_m^2\Pi({\bf q},\omega+i0^+)\right] 
\Big\}.
\end{aligned}
\label{EqIntB}
\end{equation}
Here, the integration over the first term within the brace will cancel with $\Omega_0^{\rm M}$ in the total thermodynamic potential. Then, the total thermodynamic potential is given as $\Omega=\Omega_0^{\rm B}+\tilde{\Omega}_{\rm int}$ where $\tilde{\Omega}_{\rm int}$ is the integration of only the second term left in the brace in Eq.~\eqref{EqIntB}.
After substituting the bare parameters according to Eq.\eqref{EqNorm}, $\tilde{\Omega}_{\rm int}$ can be simply  written as 
\begin{equation}
\tilde{\Omega}_{\rm int}=\sum_{m,{\bf q}} \int_{-\infty}^{+\infty}\frac{d \omega}{\pi}\frac{1}{e^{\beta\omega}-1}\tilde{\delta}^d_m({\bf q}, z),
\end{equation}
in which phase $\tilde{\delta}^d_m$ is defined as
\begin{equation}
\begin{aligned}
\tilde{\delta}^d_m({\bf q},z)=\text{Arg} \Big[
\frac{M}{4\pi}D_m^{-1}+z\frac{M^2}{4\pi}v_m^{-1}+z^2\frac{M^3}{4\pi}R_m^{-1}\\
+\Pi^d_r({\bf q},z+i0^+)
\Big],
\end{aligned}
\label{Eqdelta}
\end{equation}
where $z\equiv \omega-q^2/4M+2\mu$ is the energy shifted with respect to the dimer energy. Within $\tilde{\delta}^d_m$, the regularized pair-pair propagator is given by 
\begin{equation}
\begin{aligned}
\Pi^d_r({\bf q},z)=
&-\frac{M}{V}\sum_{\bf k}k^2
-z\frac{M^2}{V}\sum_{\bf k}1 
-z^2\frac{M^3}{V}\sum_{\bf k}k^{-2}\\
&+\Pi^d({\bf q},z),
\end{aligned}
\end{equation}
where the first three terms will cancel the divergence within $\Pi^d({\bf q},z)$. At high temperature limit, $\Pi^d_r({\bf q},z)$ can be carried out explicitly as $\Pi^d_r({\bf q},z)\to-{M^{7/2}}z^3/4\pi\sqrt{-z}$ which is real for negative $z$ and pure imaginary for positive $z$. The pole of the real part of $\tilde{\delta}^d_m({\bf q},z)$ at low energy $E^d_b\simeq-v_m/MD_m$ gives the shallow bound state energy with positive $D_m$ since $v_m$ is always positive for $d$-wave resonance. The resultant deep bound state, with energy beyond the low energy expansion assumption, is associated with the ghost field with negative probability. Thus we will only keep the first shallow bound state during our calculation.

For the $p$-wave resonance, a standard NSR calculation leads to a similar total thermodynamic potential $\Omega=\Omega_0^{\rm B}+\tilde{\Omega}_{\rm int}$ with 
\begin{equation}
\begin{aligned}
\tilde{\delta}_m^p({\bf q},z)=\text{Arg}\Big[\frac{M}{4\pi}v_m^{-1}+z\frac{M^2}{4\pi}R_m^{-1}
+\Pi^p_r({\bf q},z+i0^+)\Big] ,
\end{aligned}
\end{equation}
where $\Pi^p_r({\bf q},z+i0^+)$ is the regularized pair-pair propagation:
\begin{equation}
\begin{aligned}
\Pi^p_r({\bf q},z)=-\frac{M}{V}\sum_{{\bf k}}1-z\frac{M^2}{ V}\sum_{{\bf k}}\frac{1}{k^2}+\Pi^{l=1}({\bf q},z),
\end{aligned}
\end{equation}
which gives a shallow bound state at $E_b^p\simeq-R_m/Mv_m$ at strong coupling regime with positive $v_m$.

\section{The equation of states} For the high-partial wave resonance, the summation over $m$ takes $0, \pm1, ..., \pm l$. Due to the splitting of the resonances, it will be safe to only take $m=0$ channel in our consideration~\cite{dwaveExp}. For multiply channel cases, qualitative behavior will be similar according to Ref.~\cite{Juan2016}. 

We study the equation of state of the system through the free energy $F=\Omega+\mu N$, where the chemical potential is determined according to the number equation $N=-\frac{\partial \Omega}{\partial \mu}$.  Generally the free energy can be written as a universal form of $F=NE_nf(k_F=nR_d,k_n^3v_d,k_n^5D_d)$ or $F=NE_nf(k_nR_p,k_n^3v_p)$ for the $d$- or $p$-wave resonance. Here energy $E_n=k_n^2/2M$ and $k_n$ is Fermi momentum defined through the total number of particles. Note that $E_n$ of a single and two components Bose gas are different with the same total number of particles $N$.
Instead of $\{R_d, v_d, D_d\}$ or $\{R_p, v_p\}$, a physical parameter, the ratio of the approximated binding energy $-E_b$ and $E_n$ is chosen to characterize the coupling strength. 
\begin{figure}[t]
\includegraphics[width=0.45\textwidth]{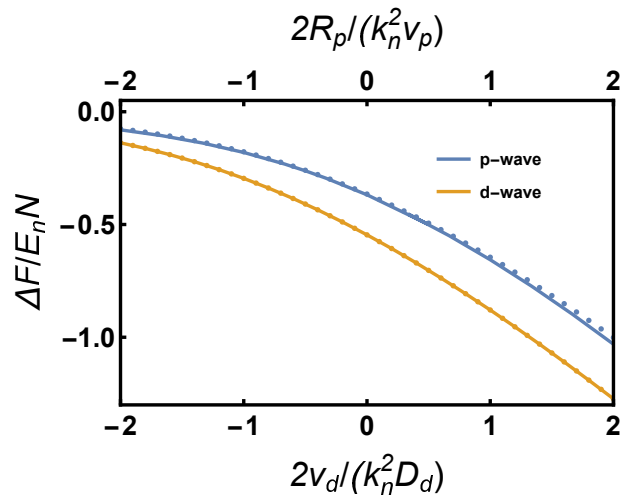}
\caption{When $k_{\rm B}T=E_n$, the solid lines are the relative free energy $\Delta F=F-F_0$ as a function of the ratio between binding energy and $E_n$ within NSR calculation. During the calculation, $1/k_nR_d=1/k_nR_p=30$ and $1/k_n^3v_d=5\times10^4$. The dots stand for the relative free energy as a function of detuning over $E_n$, which is calculated from free two channel model in terms of single ($d$-wave) and two ($p$-wave) component Bose system.}
\label{Fig1}
\end{figure}
We plot $\Delta F=F-F_0$ in Fig.~\ref{Fig1} (solid lines) for both $p$- and $d$-wave resonances. 
In general, $\Delta F$ monotonically decreases in terms of stronger coupling strength. 
Both $p$- and $d$-wave resonance can be sketched by a simple free two channel model:
\begin{equation}
\hat{H}=\sum_{{\bf k},\sigma}\left(\frac{k^2}{2M}-\mu\right) a^\dagger_{{\bf k},\sigma}a_{{\bf k},\sigma}+\sum_{\bf q}\left(\frac{q^2}{4M}-\bar{\nu}-2\mu\right)b^\dagger_{{\bf k}}b_{{\bf k}},
\end{equation}
where no coupling term is presented. The spin component $\sigma$ will be chosen according to the type of resonance. Within this free model, the resultant free energy as a function of detuning $\bar{\nu}$ is plotted as the dots in Fig.~\ref{Fig1} which overlap the solid lines. Thus as long as $E_b/E_n$ is fixed, the equation of state is determined even with difference values of $\{k_nR_d,k_n^3v_d,k_n^5D_d\}$ or $\{k_nR_p,k_n^3v_p\}$. This is the underline physical reason why the $-E_b/E_n$ ($2v_d/k_n^2D_d$ or $2R_d/k_n^2v_p$) can be chosen as the universal parameter characterizing the coupling strength.

\section{Critical temperature of the condensation} When temperature is low enough, the celebrated Bose-Einstein condensation will occur for a Bose gas. Meanwhile, attractive particles can also condense in the form of pairs. A natural question is raised: how these two kinds of condensation compete with each other through the corssover? We assume the former happens at $\mu=0$ and determine the latter critical temperature according to the Thouless criterion. 
\begin{figure}[t]
\includegraphics[width=0.45\textwidth]{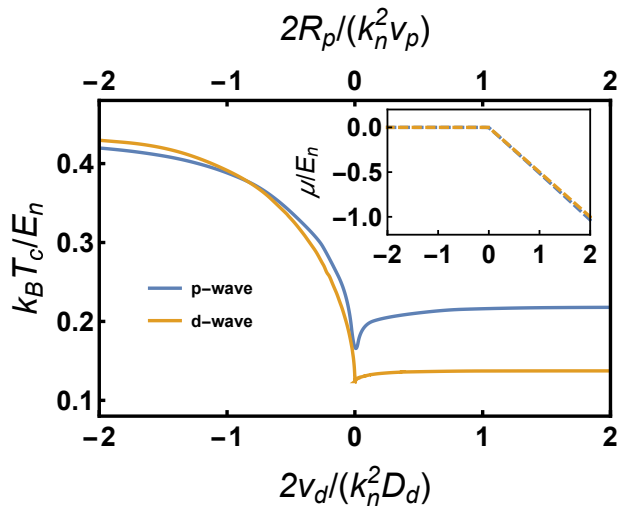}
\caption{The critical temperature $T_c$ from the weak to strong coupling limit. Inset: the corresponding chemical potential at each coupling strength.
At weak coupling side, system will condense in the form of solitary atoms at $T_c$ with $\mu=0$. While at strong coupling side, it condenses in the form of molecules with binding energy given by $|2\mu|$.}
\label{Fig2}
\end{figure}

Within the NSR method, the critical temperature $T_c$ from the weak to strong coupling is shown in Fig.~\ref{Fig2}.

In the weak coupling limit, $T_c$ reaches the same asymptotic value $0.436E_n$ for the two high partial wave resonances, which is also the critical temperature of condensation of $N$ free bosonic atoms. In this limit, system can be viewed as free Bose gas and it is reasonable to get a solitary atomic condensation with $\mu=0$. System obtain a atomic superfluid phase. However, for $s$-wave resonance, in the weak coupling side, the condensation of atom is always accompanied by the condensation of pairs due to the coupling between condensed atoms and pairs~\cite{Radzihovsky2004, Koetsier2009}. However, for the high partial wave case, the condensed atoms at zero momentum is decoupled to the pairs. Thus a pure atomic superfluid phase is obtained.  In the strong coupling limit, $T_c$ approaches $0.137E_n$ and $0.218E_n$ respectively for the $d$- and $p$-wave interaction. Although the two asymptotic values appear different, both of them are the condensation temperature of $N/2$ particle with mass $2M$. The difference is due to the definition of Fermi energy $E_n$ for single and two components Bose system with the same $N$.
In the strong coupling regime, a condensation of molecules occurs with bound state energy given by $2\mu$ (inset of Fig.~\ref{Fig2}). Through the crossover, a phase transition between atomic superfluid phase and molecular superfluid phase is achieved. The lowest $T_c$ occurs around unitary point where atomic superfluid and molecular superfluid coexist and meet the normal phase 
at the multi-critical point.

\section{Particle population distributions} With respect to the two terms within thermodynamic potential, the total number of particles $N$ also contains two parts:
\begin{equation}
N=N_{\rm Atom}+N_{\rm Dimer},
\end{equation}
with $N_{\rm Atom}=-\frac{\partial\Omega_0^{\rm B}}{\partial \mu}$ and $N_{\rm Dimer}=-\frac{\partial\tilde{\Omega}_{\rm int}}{\partial \mu}$. Here we view $N_{\rm Atom}$ is related to the number of particles without interaction that remains the form of solitary atoms while $N_{\rm Dimer}$ is associated with the number of particles which form dimers due to attractive interactions. 
\begin{figure}[t]
\includegraphics[width=0.45\textwidth]{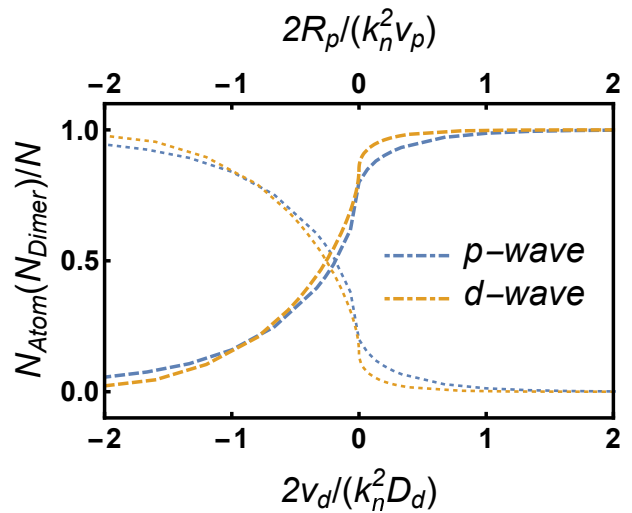}
\caption{Through the crossover, the atomic $N_{\rm Atom}$ and dimer fraction $N_{\rm Dimer}$ distribution at critical temperature $T_c$. The dotted lines are the atomic fraction $N_{Atom}/N$ and dashed lines are the dimer fraction $N_{\rm Dimer}/N$ with different colors corresponding to different type of partial wave. }
\label{Fig3}
\end{figure}

\begin{figure}[t]
\includegraphics[width=0.45\textwidth]{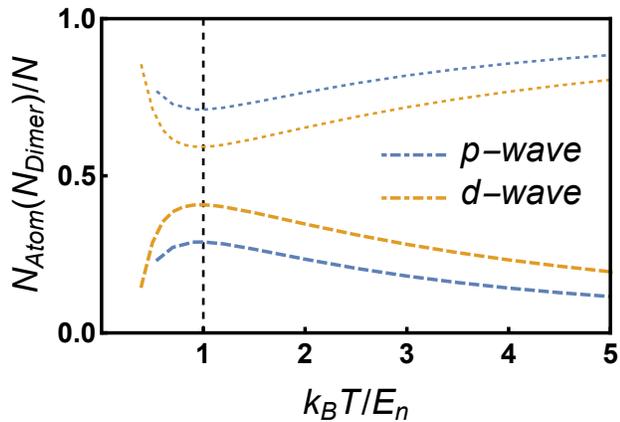}
\caption{When $2v_d/k_n^2D_d=2R_p/k_n^2v_p=-1$ in the weak coupling side, plotting the particle populations in terms of $T$. The dotted lines are the atomic fraction $N_{Atom}/N$ and dashed lines are the dimer fraction $N_{\rm Dimer}/N$ with different colors corresponding to different type of partial wave. The black dashed line corresponds to $k_BT=E_n$.}
\label{Fig4}
\end{figure}

Through the crossover we extract $N_{{\rm Atom}({\rm Dimer})}$ at each corresponding critical temperature $T_c$ (Fig.~\ref{Fig2}). 
As shown in the Fig~\ref{Fig3}, in the weak coupling limit,  the atomic fraction approaches unit and dimer fraction approximates zero. The system can be regarded as noninteracting free solitary atoms. It is reasonable to get $100\%$ atomic fraction.
In the strong coupling limit, fraction of atoms and dimers is inverse. System energy is dominated by the dimer bound state so that almost all particles formed dimers in this limit. 
In between, atomic fraction and dimer fraction compete with each other and two components coexist. 
As shown in the Fig~\ref{Fig3},  in two limits, particle population distribution is independent on the resonance type. The conversion from atoms to dimers seems much faster for $d$-wave interaction with $d$-wave dimer fraction saturates around $-E_b/E_n=0.5$ and the $p$-wave saturates around $-E_b/E_n=1$. This is because the $d$-wave resonance obtains a lower $k_BT_c/E_n$ in the strong coupling regime (Fig.\ref{Fig2}).  Generally a low temperature will benefit the formation of dimers. This argument is valid in the strong coupling side with existence of the bound state. However, in the weak coupling side, a modification to the above argument  is needed as explained as follows. 
In Fig.~\ref{Fig4}, we provide the particle fraction in terms of temperature when $2v_d/k_n^2D_d=2R_p/k_n^2v_p=-1$.
Now the bound state energy $E_b=E_n>0$ and a quasi-bound state exist due to the centrifugal barrier for high partial wave interactions. As shown in Fig.~\ref{Fig4}, generally, the dimer fraction increases at lower temperature. However, a maximum (minimum) of dimer (atom) fraction appears around $k_BT=E_n$. A turning point presents in the atom and dimer fraction distribution. Around the turning point, coincidently we also have $k_BT=E_b$ with quasi-bound state energy approximates the temperature. Thus we can claim that the nonmonotonic behavior of the particle distribution is due to a resonance conversion between atoms and dimers. Namely, when $k_BT=E_b$, dimers constrained in the centrifugal barrier are resonantly coupled free atoms outside, which manifests as the turning points in the particle population distributions in Fig.~\ref{Fig4}.  

\section{Conclusion}
Within NSR approximation, we show that in the week coupling limit, high partial wave Bose gas will condense with solitary atoms. In the strong coupling limit, the system will condense in the form of pairs. In terms of coupling  strength, a transition between atomic superfluid  and molecular superfluid is achieved. At the lowest $T_c$, normal phase, atomic superfluid phase and molecular superfluid phase meet at the multi-critical point. 
Finally, an investigation of the effect of coupling strength and temperature  on the form of dimers is given. Stronger interaction will enhance the formation of dimers. In general, low temperature benefits the dimer formation. However, when $k_BT=E_b$, a resonant conversion between atoms and dimers is identified through the appearance of the turning points in the particle population distribution. For the experiment, the most difficulties in manipulating the Feshbach resonance lie in the strong coupling side or near the resonance. Since the resonance conversion identified here occurs in the weakly coupling side, it seems very promising to observe the turning point indicating the resonance conversion upon the realization of high partial wave Feshbach resonances. 

{\em Acknowledgement}. We thank Hui Zhai for helpful discussion. Ran Qi is supported by NSFC under Grants No. 11774426, and the Fundamental Research Funds for the
Central Universities and the Research Funds of Renmin University of China under Grants No. 15XNLF18 and No. 16XNLQ03.


\begin{thebibliography}{99}
\bibitem{Luciuk2016}
C. Luciuk, S. Trotzky, S. Smale, Z. Yu, S. Zhang, and J. H. Thywissen, Nat. Phys. \textbf{12}, 1 (2016).
\bibitem{Yu2015}
Z. Yu, J. H. Thywissen, and S. Zhang, Phys. Rev. Lett. \textbf{115}, 135304 (2015).
\bibitem{Yoshida2015}
S. M. Yoshida and M. Ueda, Phys. Rev. Lett. \textbf{115}, 135303 (2015).
\bibitem{Mingyuan2016}
Mingyuan He, Shaoliang Zhang, Hon Ming Chan, and Qi Zhou, Phys. Rev. Lett. \textbf{116}, 045301 (2016).
\bibitem{Cui2016}
X. Cui and H. Dong, Phys. Rev. A \textbf{94}, 063650 (2016).
\bibitem{Peng2016a}
S.-G. Peng, X.-J. Liu, and H. Hu, Phys. Rev. A \textbf{94}, 063651 (2016).
\bibitem{Regal2003}  
C. A. Regal, C. Ticknor, J. L. Bohn, and D. S. Jin, Phys. Rev. Lett. \textbf{90}, 053201 (2003).

\bibitem{Zhang2004}
J. Zhang, E. G. M. van Kempen, T. Bourdel, L. Khaykovich, J. Cubizolles, F. Chevy, M. Teichmann, L. Tarruell, S. J. J. M. F. Kokkelmans, and C. Salomon, Phys. Rev. A \textbf{70}, 030702 (2004).
\bibitem{Gunter2005} 
K. Gunter, T. Stoferle, H. Moritz, M. Kohl, and T. Esslinger, Phys. Rev. Lett. \textbf{95}, 230401 (2005).
\bibitem{Schunck2005} 
C. H. Schunck, M. W. Zwierlein, C. A. Stan, S. M. F. Raupach, W. Ketterle, A. Simoni, E. Tiesinga, C. J. Williams, and P. S. Julienne, Phys. Rev. A \textbf{71}, 045601 (2005).
\bibitem{Rem2013} B. S. Rem, A. T. Grier, I. Ferrier-Barbut, U. Eismann, T. Langen, N. Navon, L. Khaykovich, F. Werner, D. S. Petrov, F. Chevy, and C. Salomon, Phys. Rev. Lett. \textbf{110}, 163202 (2013).
\bibitem{Stoof1994} H. T. C. Stoof, Phys. Rev. A \textbf{49}, 3824 (1994).
\bibitem{dwaveExp}
Xing-Can Yao, Ran Qi, Xiang-Pei Liu, Xiao-Qiong Wang, Yu-Xuan Wang, Yu-Ping Wu, Hao-Ze Chen, Peng Zhang, Hui Zhai, Yu-Ao Chen, Jian-Wei Pan, arXiv:1711.06622 (2017).
\bibitem{Cui2017}
Y. Cui, C. Shen, M. Deng, S. Dong, C. Chen, R. Lu, B. Gao, M. K. Tey, and L. You, Phys. Rev. Lett. \textbf{119}, 203402 (2017).
\bibitem{Dong2016a}
S. Dong, Y. Cui, C. Shen, Y. Wu, M. K. Tey, L. You, and B. Gao, Phys. Rev. A \textbf{94}, 062702 (2016).
\bibitem{Pengfei2016}
Pengfei Zhang, Shizhong Zhang and Zhenhua Yu, Phys. Rev. A  \textbf{95}, 043609 (2017).
\bibitem{Nozires1985}
P. Nozi{\`e}res and S. Schmitt-Rink, J. Low Temp. Phys. \textbf{59}, 195 (1985).
\bibitem{Juan2016}
Juan Yao and Shizhong Zhang, arXiv:1609.06476 (2016).
\bibitem{Radzihovsky2004}
L. Radzihovsky, J. Park, and P. B. Weichman, Phys. Rev. Lett. \textbf{92}, 160402 (2004).
\bibitem{Koetsier2009}
A. Koetsier, P. Massignan, R. A. Duine, and H. T. C. Stoof, Phys. Rev. A \textbf{79}, 063609 (2009).
\end{thebibliography}
\end{document}